# Review and Analysis of Pain Research Literature through Keyword Co-occurrence Networks

**Burcu Ozek, Zhenyuan Lu, Fatemeh Pouromran, Sagar Kamarthi**

Mechanical and Industrial Engineering Department, Northeastern University,

Boston, Massachusetts, United States of America

## Abstract

Pain is a significant public health problem as the number of individuals with a history of pain globally keeps growing. In response, many synergistic research areas have been coming together to address pain-related issues. This work conducts a review and analysis of a vast body of pain-related literature using the keyword co-occurrence network (KCN) methodology. In this method, a set of KCNs is constructed by treating keywords as nodes and the co-occurrence of keywords as links between the nodes. Since keywords represent the knowledge components of research articles, analysis of KCNs will reveal the knowledge structure and research trends in the literature. This study extracted and analyzed keywords from 264,560 pain-related research articles indexed in IEEE, PubMed, Engineering Village, and Web of Science published between 2002 and 2021. We observed rapid growth in pain literature in the last two decades: the number of articles has grown nearly threefold, and the number of keywords has grown by a factor of 7. We identified emerging and declining research trends in sensors/methods, biomedical, and treatment tracks. We also extracted the most frequently co-occurring keyword pairs and clusters to help researchers recognize the synergies among different pain-related topics.

Sagar Kamarthi: s.kamarthi@northeastern.edu

1. Introduction

Pain is an uncomfortable sensory and emotional experience, which serves as a symptom of various medical conditions[1–3]. Pain occurs due to multiple causes (e.g., broken bone, strained muscle, or surgery), at different body locations (e.g., back pain, muscle pain, knee pain, or chest pain), and in various forms (e.g., acute pain, chronic pain, or intermittent pain) [3]. In the United States, 20.4% of adults have chronic pain, and 7.4% of adults report that their lives are significantly impacted by chronic pain, according to the 2019 National Health Interview Survey [4,5]. Health economists estimated that the annual cost of chronic pain in the USA is around $635 billion, which exceeds the yearly cost due to diabetes ($188 billion), cancer ($243 billion), or heart disease ($309 billion) [6].

Recognizing pain as a public health problem, researchers are rapidly expanding pain-related studies across different synergistic fields. Since 2002, roughly 291,560 pain-related articles have been published in the scientific literature indexed in IEEE, Web of Science, Engineering Village, and PubMed. It is important to review the literature to assist researchers cognizant of existing knowledge, identify knowledge gaps, advance research in an impactful direction, and generate new knowledge [7]. However, a manual review of the vast amount of literature is complex and time-consuming. A promising and viable approach is to employ an automated literature review method to gain insights from the vast literature. Network methodologies are helpful in providing an efficient, high-level, and automatic literature review process to the researchers. The network analysis can reveal hidden patterns and relationships between interconnected and interrelated components of a large-scale complex system [8]. The aim of the network methodology for the literature review process is to extract meaningful information from the underlying literature, provide knowledge maps and structures, and discover research trends using various literature components such as authors, institutes, citations, or keywords [9]. Network-based methods are generally called bibliometric networks. The following are the most studied bibliometric networks: (1) Collaboration Networks, (2) Citation Networks, and (3) Keyword Co-occurrence Networks (KCNs) [10–12]. For the automatic literature review, KCNs are more apt than collaboration and citation networks because KCNs reveal the relations among knowledge elements, the relative importance of knowledge elements, emerging topics, and the evolution of the subject over time [9,13].

This study uses KCNs to review and analyze "pain" literature automatically. It analyzes various network parameters considering centrality, affinity, and cohesiveness among the keywords to provide a knowledge map of pain research. To build KCNs, we extracted keywords from 264,560 articles indexed in IEEE, PubMed, Engineering Village, and Web of Science between 2002 and 2021. By applying text mining techniques, we eliminated irrelevant and redundant keywords. We classified keywords into three tracks: sensors/methods-related keywords (e.g., electromyography, biomarker, machine learning), biomedical-related keywords (e.g., chronic pain, back pain, acute pain), and treatment-related keywords (e.g., surgery, acupuncture, medication) to organize the literature review for easy understanding. To the best of our knowledge, no automated literature review method was employed to gain insights from the vast amount of pain-related literature.

This paper is organized as follows. The background section explores the pain-related review papers published in the literature. The methods section presents an overview of the KCN-based approach to review and explore an extensive amount of literature. It also describes the data collection and data preprocessing tasks employed to build KCNs. The results and discussion section identifies and describes the emerging topics and their implications in pain research. The conclusion section summarizes the findings and limitations of this work and comments on the future direction of this work.

2. Background

This section reviews existing studies on pain management research from various angles. The existing literature review studies broadly fall into two categories: (1) objective pain assessment trends and (2) physiological mechanisms and treatment trends.

*Objective pain assessment trends*: The articles have explored automatic pain assessment [14–16]. In current clinical settings, patients typically self-report their pain level using the Verbal Rating Scale (VRS), Visual Analogue Scale (VAS), and Numerical Rating Scale (NRS)[17,18]. These self-reported pain measurements are subjective and often impractical when patients are not in an alert state or unable to communicate their pain level. Infants, toddlers, and adults with communication or cognitive deficits may not be able to convey their pain levels verbally. Researchers have explored the physiological signals and behavioral responses for objective pain measurement [2,19–23]. Werner et al.[24] presented a survey of automated pain recognition-related papers indexed in the Web of Science. They emphasized the advancements in non-contact and contact-based automatic pain recognition techniques that use facial expression, voice, physiology, and multimodal information. Lötsch et al.[25] published a review on machine learning in pain research to raise knowledge of the approaches in ongoing and upcoming projects. Wagemakers et al.[26] provided an in-depth analysis of the devices and methods for objectively measuring patients' pain. Chen et al.[27] reviewed various wearable physiological and behavioral sensors that may help build automated monitoring systems for pain detection in clinical settings. Zamzmi et al.[28,29] reviewed features, classification tasks, and databases for automated pain assessment in infants and provided pain assessment techniques in children considering physiological and behavioral scope.

*Physiological mechanisms and treatment trends*: Several researchers have tried to understand the mechanism of pain and devise methods for better treatment of a specific type of pain [30–33]. Koechlin et al.[34] presented a systematic review of the role of emotion regulation in chronic pain. They examined the risk and protective factors contributing to chronic pain management. IsHak et al.[35] examined studies that addressed pain comorbid with depression through a systematic review. They observed that depression and pain are highly related and may worsen physical and psychological symptoms. Shraim et al.[36] conducted a systematic review and synthesized classification based on the mechanism of pain in the musculoskeletal system. They evaluated methods to distinguish three categories of pain mechanisms: nociceptive, neuropathic, and nociplastic pain [37]. Urits et al.[38] conducted an exhaustive literature review of low back pain and examined the pathophysiology, diagnosing methods, and treatment strategies. Finnerup et al.[39] reviewed neuropathic pain topics with emphasis on its mechanism and treatments.

The present study addresses the timely need for an exhaustive review of the pain literature published in the last two decades. We consider that the scientific research publications reflect the emerging trends in pain research and we employed a KCN-based method to comprehensively analyze thousands of pain-related articles indexed in IEEE, PubMed, Engineering Village, and Web of Science databases. This approach addresses the challenges posed classical literature review approaches in terms of time and effort to capture insights from a vast body of literature.

## 3. Methods

### 3.1 Overview of the Existing Methods

In literature, a few studies utilized KCNs to analyze research fields. Lee et al. [40] developed a KCN for "regional innovation systems (RIS)" literature and collected 432 articles to investigate the development of RIS research and future research directions. They used centrality-related network features. Li et al. [41] built a KCN using the complex-network-related keywords extracted from 5,944 articles published between 1990 and 2013 to analyze the trends and relationships between knowledge elements. They evaluated the networks considering the degree, clustering coefficient, and shortest path principles to understand the evolution of the articles.

Radhakrishnan et al. [42] created a novel KCN-based method to help researchers review scientific literature. As a case study, they built KCNs for "nano-related environmental, health, and safety (EHS) risk" literature. They collected keywords from 627 papers published between 2000 and 2013. They used network parameters such as degree, strength, average weight, weighted nearest neighbor's degree, and clustering coefficient for statistical analysis of KCNs.

Yuan et al. [13] presented a KCN-based analysis of the data science trends in the manufacturing literature. They extracted keywords from a collection of 84,041 articles published between 2000 and 2020. They categorized the keywords according to the nine pillars of Industry 4.0 to understand the emerging topics in this smart manufacturing research.

Weerasekara et al. [43] reviewed the evolution of industry 4.0 for asset life cycle management for sustainability concepts using KCN-based techniques. They extracted keywords from 3,896 articles and analyzed the research trends.

A common objective of the above-mentioned studies is to help researchers understand the trends in specific literature and guide them to identify future directions. In the present work, we adopted these complex-network-related metrics from Radhakrishnan et al. [42] to build algorithms for reviewing and analyzing large-scale pain research literature in the current study.

### 3.2 Data Collection and Processing Procedure

This section discusses the steps of the proposed approach to extract and analyze the keywords, which are the uncontrolled terms specified by the authors of the articles.
**Step 1**: Select the databases of citations of research articles.

**Step 2**: Develop an information-extraction procedure to collect the corpus of keywords from the research articles indexed in the selected databases.
**Step 3**: Convert the corpus into a list of unique words using the natural language processing methods.
**Step 4**: Generate adjacency matrices and weighted adjacency matrices for four-year windows: 2002 – 2006, 2007 – 2011, 2012 – 2016, and 2017 – 2021.
**Step 5**: Construct KCNs from the adjacency matrices and weighted adjacency matrices.
**Step 6**: Conduct review and analysis of pain literature using the KCNs.

**Article Search.** To create the KCNs of pain research articles, we searched for articles that included the term "pain" in the articles' titles and keywords (corpus) defined by the authors. These articles are quarried from IEEE, Engineering Village, and Web of Science with a scope of 2002 to 2021. The National Center for Biotechnology Information (NCBI) provides the Entrez system currently covering a variety of biomedical data among 38 databases, including the PubMed database. We wrote a python code to connect the Entrez system to request and retrieve our targeted data from the PubMed database. This search resulted in a total of 184,174 articles from PubMed. From the other three databases, we searched for the articles through manual queries. This resulted in 888, 6622, and 229,758 articles from IEEE, Engineering Village, and Web of Science, respectively. Then all the titles and keywords of these articles were extracted and prepared for the next step.

**Article Screening.** We removed duplicate articles within and across databases. This process eliminated 156,882 articles leaving 264,560 articles for the next stage. The majority of the redundant corpus was between Web of Science and PubMed. Approximately half of the articles from both databases were duplicates.

**Text Indexing.** The corpus of keywords from the articles is unstructured data. We processed this data to bring it into a structured data format through the following procedure (see **Figure 1**):

- Converted all keywords to lowercase to make them case-uniform, e.g., changed "Neuropathic Pain" to "neuropathic pain"
- Extracted individual keywords separated by delimiters such as ":;-/", e.g., changed "machine learning; SVM" to "machine learning" and "SVM"
- Removed punctuation marks within the keywords: !"#$%&\'()*+,./:;<=>?@[\\]^_`{|}~, e.g., changed "1,3,4 – Oxadiazole" to "134 – Oxadiazole"
- Tokenized keywords involving multiple words separated by spaces and hyphens, e.g., changed "neuropathic-pain" to "neuropathic" and "pain".
- Extracted stem words by converting every single word into its root using language rules, e.g., extracted "neuropath" from "neuropathic".
- Converted all nouns with plural forms into a singular form, e.g., removed "s", "es", or changed "children" to "child".
- Transformed the verbs in the noun form, verbs in the past tense to root words, e.g., changed words ending "ed" and "ing" to their root words.

- Changed terms with postfix, "ly", "est", "ation", or "ment" to their root form, e.g., simplified "assessment" to "assess".
- Concatenated back processed words with a blank space between words to form original keywords, e.g., concatenated "neuropath" and "pain" to form "neuropath pain". This gave us sets of keywords constructed with tokenized and stemmed words.
- Dropped all the terms of length one or two letters which, to our knowledge, did not play a crucial part in our analysis.

**Adjacency Matrix and Frequency Table**. Using the indexed corpus from the procedure described above, we generated a document frequency table and multiple adjacency matrices of co-occurring keywords, and a dictionary including those keywords before and after stemming. For example, we present the top 10 frequently used words before and after stemming in **Table 1**

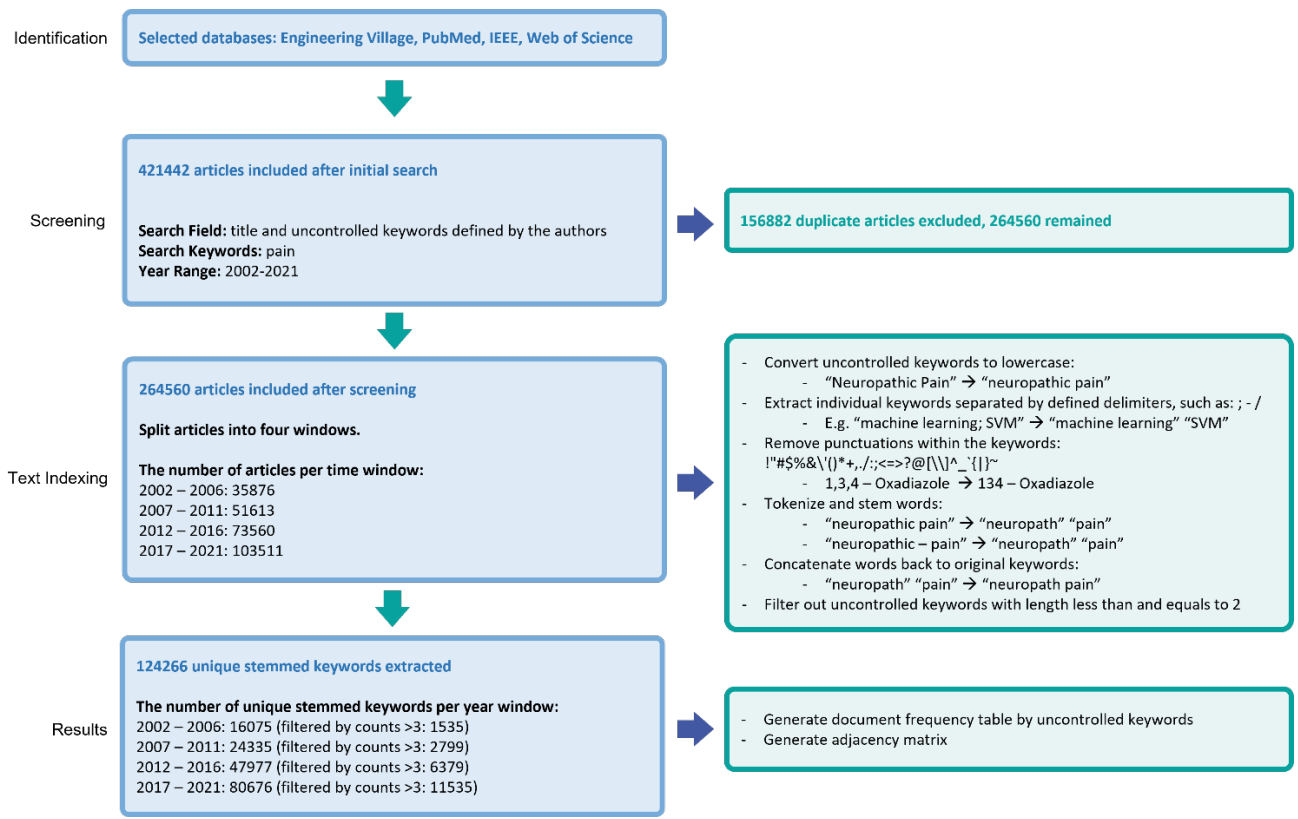

**Figure 1.** The process of data collection, processing, and cleaning by text mining techniques.

**Table 1.** The keywords before and after stemming in the latest year window (2017-2021).

| Original Keywords (before stemming) | Stemmed Keywords (after stemming) |
| --- | --- |
| painfulness; painful; pains; pained; pain | pain |
| chronic pains; chronic pain | chronic pain |
| low back pains; low back pain | low back pain |
| pain managers; pain management; pain managements | pain manag |
| neuropathic pains; neuropathic pain | neuropath pain |
| opioides; opioid; opioide; opioids | opioid |
| postop pain; postoperative pain | postop pain |
| analgesia | analgesia |
| quality life | qualiti life |
| abdominal pains; abdominal pain | abdomin pain |

### *3.3 Network*

A network is a set of connected entities. In network theory, entities in the graph are named nodes or vertices, and the connections between the entities are called links or edges [44]. Throughout the study, we use "edges" and "links" interchangeably. This study creates KCNs, which explore the knowledge structure of the body of scientific literature by investigating the relations among keywords in the field [42]. In a KCN, nodes represent the keywords collected from the articles, and the edges represent the co-occurrences between pairs of keywords. An edge connects a pair of nodes (keywords) if the keywords co-occur in an article.

**Figure 2** illustrates an example of a KCN. Nodes are the top five frequently used keywords in the literature between 2017 and 2021, namely, pain, chronic pain, low back pain, pain management, and neuropathic pain. If a pair of words are connected by a link, then these keywords co-occur or vice versa. The thickness of the link indicates the number of times the keywords co-occur in the pool of articles: the heavier the link, the higher the co-occurrence counts. For example, pain and chronic pain co-occurred 290 times, but low back pain and neuropathic pain co-occurred only 19 times.

In this study, the relationship between keywords does not have a direction. A link simply represents that keyword $i$ co-occurs with keyword $j$. Therefore, a KCN is an undirected network. In addition to that, it is a weighted network since the connections among nodes have weights assigned to them, which are the number of times the keyword-pairs co-occur.

A network can be represented as an adjacency matrix; it is a symmetric matrix with a size of $n \times n$, where $n$ is the number of nodes (i.e., keywords) [44]. In the adjacency matrix, if keywords $i$ and $j$ co-occur in an article, $a_{ij} = 1$, otherwise $a_{ij} = 0$ and the diagonal elements are assigned zero [44]. A KCN network can be treated as a weighted network by adding weights to links; in that case, the adjacency matrix becomes the weighted adjacency matrix, in which $w_{ij}$ denotes the weight of the link connecting nodes $i$ and $j$; in other words, $w_{ij}$ indicates the co-occurrence frequency of nodes (keywords) $i$ and $j$ [42]. Since a KCN is undirected, $w_{ij} = w_{ji}$.

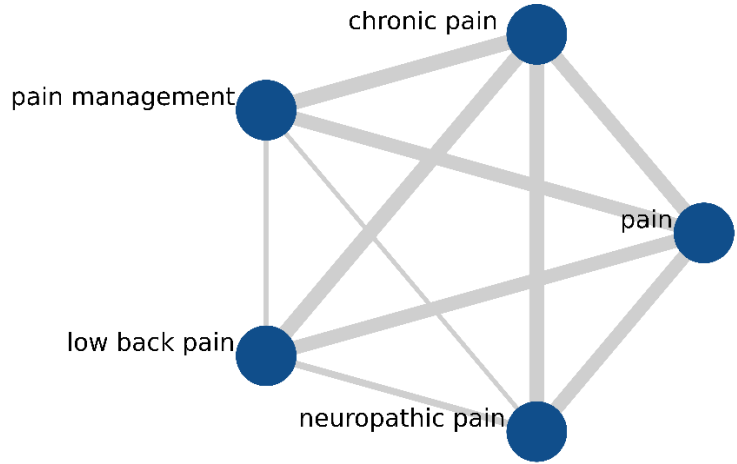

**Figure 2.** The KCN for the five most frequently used keyword during 2017-2021. Nodes (entities) are the keywords of the articles, edges (links) are the co-occurrences of pairs of keywords, and the link thickness denotes the number of times the keywords co-occur in the pool of articles.

*3.4 Network Parameters*

This study aims to investigate the KCN considering the nodes' and edges' statistical features, centrality, affinity, and cohesiveness.

**(1) Degree:**
A node's degree is the total number of direct links the node has with the other nodes [45]. It is a centrality metric and is used to measure a node's importance in graph [46]. The degree of node $i$ is defined as follows:

$$k_i = \sum_{j \in V_i} a_{ij} \qquad (1)$$

where $V_i$ represents the set of nodes connected to node $i$ and $a_{ij}$ denotes the element of the adjacency matrix indicating the presence or absence of the connection between node $i$ and node $j$.

An intuitive inference about the degree is that nodes with more connections are more central to the network. However, in a weighted graph, this is not always the case. In addition to the links, the weights of the links are to be considered.

**(2) Strength:**

A node's strength is the sum of the weights of all links connected to the node. It indicates the importance of a node considering both degree and weight [47]. The node's strength of a node $i$ is calculated as follows:

$$s_i = \sum_{j \in V_i} w_{ij} \quad (2)$$

where $V_i$ is the set of nodes connected to node $i$; $w_{ij}$ is the weight of the link between nodes $i$ and $j$ in the weighted adjacency matrix.

**(3) Average weight as a function of endpoint degree:**
The average weight as a function of endpoint degree indicates changes in the frequency of co-occurrence of the edges between pairs of nodes as the product of the degrees of end nodes of edges changes[13,42]. It determines the relative change in the edge weights as the number of edges connected to the end nodes changes [48]. The endpoint degree of an edge between node $i$ and node $j$ is defined as the product of degrees of nodes connected to the edge, i.e., $k_i k_j$. Let $Q_{ij}$ be the set weights of all edges whose endpoint degree is equal to $k_i k_j$.

$Q_{ij} = \{w_{ab} |\ k_a k_b = k_i k_j; a = 1, 2, \ldots, n; b = 1, 2, \ldots, n\}$

where and $n$ is the total number of nodes in the network. The average weight $\langle w_{ij} \rangle$ is defined as the average of all weights $w_{ab} \in Q_{ij}$.

$$\langle w_{ij} \rangle = \frac{\sum_{w_{ab} \in Q_{ij}} w_{ab}}{|Q_{ij}|} \quad (3)$$

where, $|Q_{ij}|$ is the cardinality of the set. Here is an example how to implement the **Eq. 3**:

**Step 1**: Find degree of each node
　　e.g., $k_1 = 1, k_2 = 60, k_3 = 20, k_4 = 3$
**Step 2**: Find weight of each edge
　　e.g., $w_{12} = 30; w_{13} = 35; w_{14} = 40; w_{23} = 50; w_{24} = 25; w_{34} = 100;$
**Step 3**: Find the end degree of each edge $e_{ij}$
　　e.g., $k_1 k_2 = 60; k_1 k_3 = 20; k_1 k_4 = 3; k_2 k_3 = 120; k_2 k_4 = 180; k_3 k_4 = 60;$
**Step 4**: Consider a node pair $i$ and $j$ and compute its end degree
　　e.g., $k_1 k_2 = 60$
**Step 5**: Find the weights of all the node-pairs which have the same end degree as $k_i k_j$ (e.g., $k_1 k_2$)
　　e.g., $k_1 k_2 = 60; w_{12} = 30;$
　　　　$k_3 k_4 = 60; w_{34} = 100;$
**Step 6**: Take the average of weights of node-pairs which have the same end degree as $k_i k_j$ (e.g., $k_1 k_2$)
　　e.g., $<w_{12}> \frac{30+100}{2} = 65$
**Step 7**: Apply Step 4- 6 for each pair of nodes $i$ and $j$ in the network

The average weight as a function of end point degree explains the nature of association between nodes of different degrees. If $<w_{ij}>$ increases with $k_i k_j$, links among the keywords with high-degree are more prevalent than the links among the keywords with low-degree. Conversely, if $<w_{ij}>$ reduces with $k_i k_j$, links between the keywords with low-degree are more prevalent than links between the keywords with high-degree.

### (4) Average weighted nearest neighbor's degree as a function of the degree:

The average weighted nearest neighbor's degree as a function of the degree measures the affinity among a node's direct neighbors. It demonstrates if a node in the network has similar network characteristics as its neighbors in terms of degree. An increasing trend in this function implies that nodes with high-degree are prone to bind to other nodes with high-degree, in which case the network has assortative behavior. On the other hand, a decreasing trend shows that nodes with high-degree bind mostly to nodes with low-degree, in which case the network exhibits disassortative behavior [47]. It is calculated as follows:

$$k_{nn,i}^w = \frac{1}{s_i} \sum_{j \in V_i} w_{ij} k_j \quad (4)$$

where $s_i$ is the node strength, $V_i$ is the set of nodes connected to node $i$, $w_{ij}$ is the weight of the link between node $i$ and $j$, and $k_j$ is the degree of node $i$.

### (5) Weighted clustering coefficient as a function of degree:

The weighted clustering coefficient quantifies the local cohesiveness of a node; it characterizes the node's connection density to its neighbors [42,47]. In the current study, since the network is a weighted graph, the geometric average of the subgraph edge weights is used to define the clustering coefficient[49].

$$c_i = \frac{1}{k_i(k_i - 1)} \sum_{j,k \in V_i} (\hat{w}_{ij} \hat{w}_{ik} \hat{w}_{jk})^{\frac{1}{3}} \quad (5)$$

where $k_i$ is the degree of node $i$, $V_i$ is the set of nodes connected to node $i$, the maximum weight in the network $\hat{w}_{ij} = w_{ij}/\max(w)$ normalizes the weights $w_{ij}$ [50]. If the multiple nodes have the same degree, then the weighted clustering coefficient corresponding to the degree is averaged over the nodes. In other words, the multiple weighted clustering coefficients corresponding to a degree are averaged. After computing the weighted clustering coefficient using **Eq. 5**, it is plotted as a function of degree.

### 4. *Results and Discussion*

This work aims to analyze the knowledge components, structure, and research trends in the pain literature. We organized the results and discussion into four subsections: (1) Frequently used pain-related keywords, (2) KCN analysis of pain-related literature, (3) Pain research trends, and (4) Association patterns among pain-related keywords.

## 4.1. Frequently Used Pain-Related Keywords

The strength metric (see **Eq. 2**) indicates a keyword's popularity considering the total count of co-occurrences with other keywords. Using strength, **Table 2 (a)** illustrates the top 20 keywords ranked by their strength in the 2017-2021 time window. It is evident that in this time window, the researchers concentrated on pain, chronic pain, pain management, low back pain, neuropathic pain, and opioid concepts. **Table 2 (b)** demonstrates the top 20 frequently co-occurring keyword pairs. In the years 2017 through 2021, researchers were interested in synergistic topics of "pain, opioid," "pain, analgesia," "pain, anxiety," "pain, quality," and "pain, depression." Opioid and analgesia ranked positions 1 and 2 when it comes to their association with pain; in addition, opioid and analgesic co-occurred considerably as well. A similar three-way trend is observed among anxiety, depression, and pain.

**Table 2.** (a) Top 20 keywords ranked by strength (b) The top 20 frequently co-occurring pairs.

(a)

| Keyword | Strength |
|---|---|
| pain | 66282 |
| chronic pain | 22376 |
| pain management | 13413 |
| low back pain | 13037 |
| neuropathic pain | 12768 |
| opioid | 11354 |
| analgesia | 7508 |
| quality life | 6690 |
| postoperative pain | 6544 |
| depression | 6515 |
| anxiety | 5863 |
| osteoarthritis | 5158 |
| back pain | 4942 |
| inflammation | 4655 |
| abdominal pain | 4484 |
| rehabilitation | 4457 |
| neck pain | 4144 |
| analgesic | 3908 |
| acute pain | 3749 |
| systematic review | 3544 |

(b)

| Keyword Pairs | Co-occurrence |
|---|---|
| pain, opioid | 881 |
| pain, analgesia | 756 |
| pain, anxiety | 688 |
| pain, quality life | 681 |
| pain, depression | 610 |
| pain, osteoarthritis | 544 |
| depression, anxiety | 532 |
| pain, inflammation | 531 |
| chronic pain, opioid | 498 |
| pain, cancer | 442 |
| pain management, opioid | 426 |
| chronic pain, pain management | 392 |
| pain, postoperative | 360 |
| pain, analgesic | 341 |
| pain, fatigue | 335 |
| pain, pain management | 317 |
| pain, nociceptive | 306 |
| opioid, analgesic | 292 |
| chronic pain, pain | 290 |
| pain, fibromyalgia | 281 |

**Figure 3** presents the network of the top 20 keywords ranked by strength. Nodes are keywords, and edges are the co-occurrences of the pairs of keywords. Node size represents its strength; the bigger the size, the higher the strength.

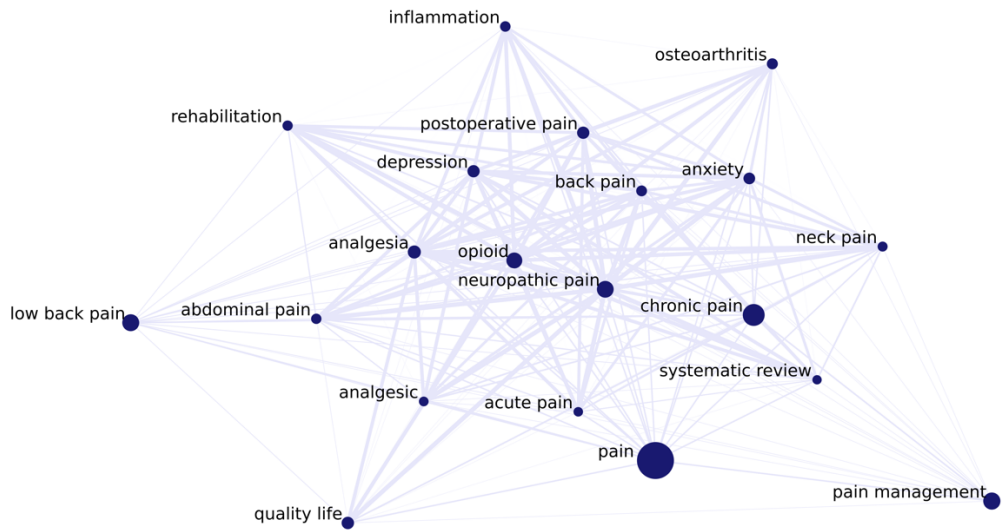

**Figure 3.** Network of the top 20 keywords ranked by strength during 2017-2021. Nodes are the keywords; edges are the co-occurrences of the pairs of keywords.

## *4.2. KCN Analysis of Pain-Related Keywords*

The summary of the topological properties of KCNs of four time-windows is presented in **Table 3** and **Figure 5**, and **Figure 5**.

From **Figure 5**, we see that the number of articles published in 2017-2021 is approximately 3 times the number in 2002-2006. During the same period, the number of unique keywords grew by a factor of 7. These statistics reveal that pain literature has expanded vastly, and new concepts have proliferated considerably in the past 20 years. In addition, the number of edges increased by a factor of 17 within the last two decades, indicating the drastic proliferation of synergies between pain-related topics.

The increasing trends in the average degree and average strength confirm the introduction of new keywords to the pain literature from diverse research fields (see **Figure 5**). The maximum degree and the maximum strength show that the keyword "pain" has formed strong connections with other keywords in the pain-related literature.

Average network weight is calculated as the sum of the weights of all links in the network divided by the total number of links. This metric remained almost the same over the four time-windows. Slight fluctuations in the average weight are due to different growth patterns in the sum of weights and the number of links. On the other hand, maximum weight has a study of growth over time.

**Table 3.** Network statistics of KCN for four time periods: 2002–2006, 2007–2011, 2012–2017, and 2017–2021. The main finding is that pain literature has grown extensively. New concepts and new connections between concepts are introduced to the pain literature.

| Metric | 2002 - 2006 | 2007 - 2011 | 2012 - 2016 | 2017 - 2021 |
|---|---|---|---|---|
| Number of Articles | 35,876 | 51,613 | 73,560 | 103,511 |
| Number of Nodes (Keywords) | 1,534 | 2,797 | 6,377 | 11,532 |
| Number of Links (Co-occurrences) | 19,927 | 48,518 | 152,808 | 346,266 |
| Average Network Degree | 25.98 | 34.69 | 47.92 | 60.05 |
| Max Degree | 1,259 | 2,187 | 4,871 | 8,407 |
| Average Network Strength | 45.40 | 56.92 | 79.80 | 106.84 |
| Max Strength | 7,706 | 13,607 | 31,918 | 66,282 |
| Average Network Weight | 1.75 | 1.64 | 1.67 | 1.78 |
| Max Weight | 165 | 211 | 393 | 881 |

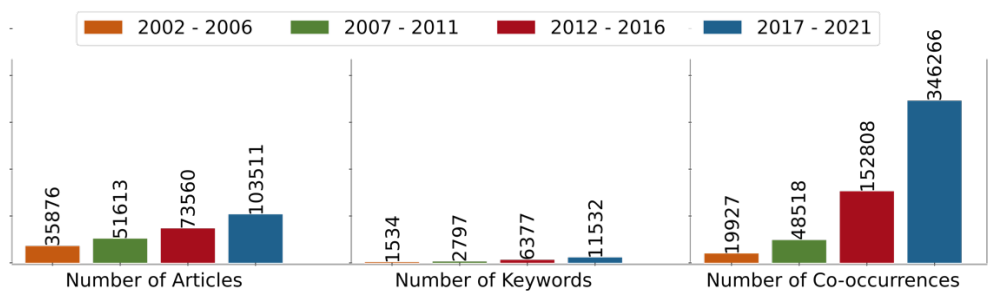

**Figure 4.** Number of articles, number of keywords, and number of co-occurrences over the four time-windows.

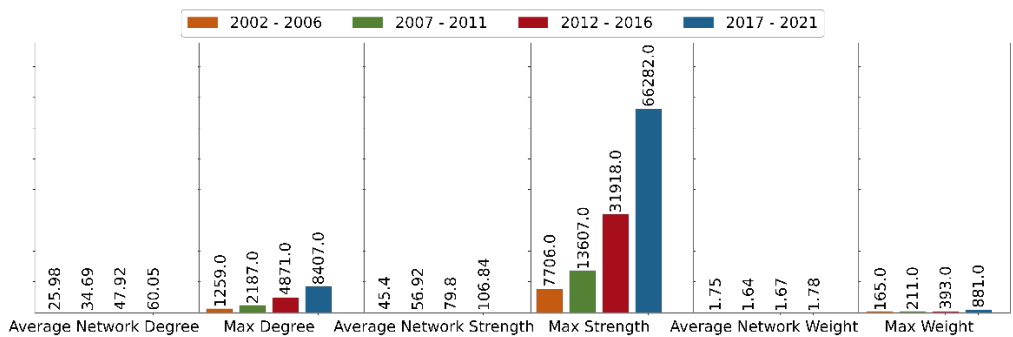

**Figure 5.** Node degree, node strength, and link weights over the four time-windows.

**Figure 6 (a)** and **(b)** show the boxplots of degree and strength. These distributions indicate the emergence of some specific keywords/topics into prominence over time. The first and the third quartile of the number of associations a keyword formed with other keywords are 16 and 52, with a median of 20. The first and the third quartile of the number of co-occurrences between pairs of keywords are 32 and 72, with a median of 19. Some extreme cases include pain management with 3264 associations with other keywords and 13413 co-occurrences in total in the last year window. A close examination of the changes in the maximum weight reveals that "Pain-Analgesia" is the most frequently co-occurring keyword pair in the first three-time periods, while "Pain-Opioid" pair dominated the 2017-2021 period.

**Figure 6 (c)** illustrates the keywords' weights distribution. The skewed distribution reveals that the cross-fertilization of topics has become extensive in recent years (2017-2021). In addition, many new keywords were introduced to the literature, which were listed only 1 or 2 times in the literature and did not form many associations with other keywords. This is the reason why the median of the weight distribution is 1.

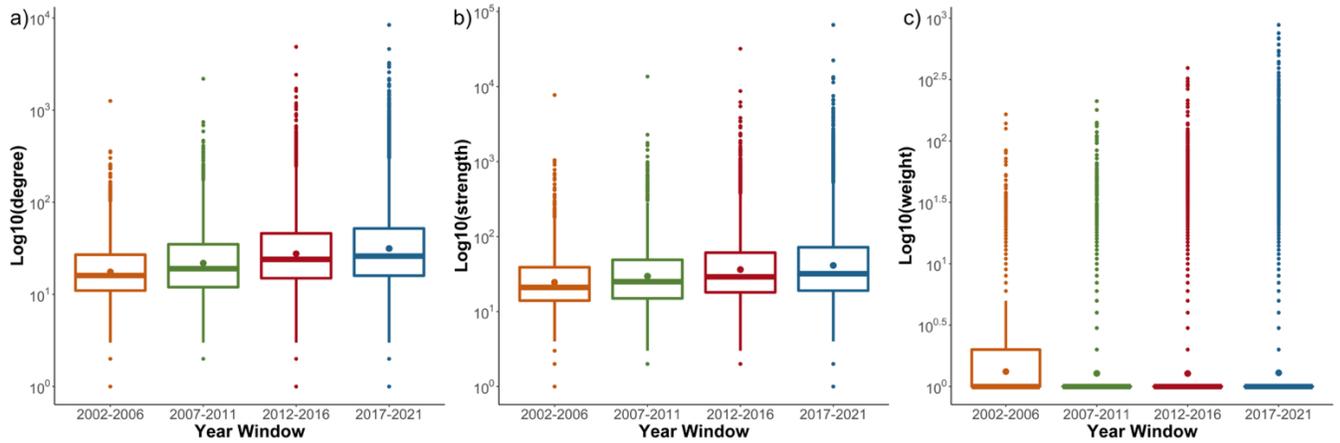

**Figure 6.** The distribution of the KCN **(a)** degree, **(b)** strength, and **(c)** weight.

The "average weight as a function of endpoint degree" is shown in **Figure 7 (a)**. It shows an increasing trend with time. The increasing trend for this metric indicates that high-degree keywords are more likely to co-occur with other high-degree keywords. Additional network properties need to be examined in parallel to confirm this conclusion. The high value for $k_i k_j$ (degree of node $i$ times degree of node $j$) may have resulted from two different possibilities: the multiplication of two high-degree nodes (e.g. $k_i k_j = 100*100 = 10,000$ ) or  the multiplication of one ultra-high-degree node and one low-degree node (e.g. $k_i k_j = 10,000*1 = 10,000$). The first case would happen when researchers frequently use the same pair of keywords (e.g., pain and opioid, pain and analgesia). The second case would happen when researchers synergize a highly popular topic with an emerging topic (e.g., pain and big data, pain and neural networks). The association between high and low-degree nodes needs to be checked to differentiate between these two cases. Therefore, the "average weighted nearest neighbor's degree" metric is examined to understand if a node and its neighbors have similar network characteristics in terms of degree.

**Figure 7 (b)** shows the "average weighted nearest neighbor's degree" vs. "degree." The time windows 2012-2016 and 2017-2021 have a slight rising trend. On the other hand, there is no such upward trend in the time windows 2002-2006 and 2007-2011; the trend is flat, but the average weighted nearest neighbor's degree fluctuates with the degree. This flat trend suggests the absence of any significant topological relationship between the "average weighted nearest neighbor's degree" and "degree." These observations contradict the fact that high-degree nodes are more likely to bind to the other high-degree nodes, and low-degree nodes bind to the other low-degree nodes. The reason is that the "average weighted nearest neighbor's degree" demonstrates that nodes do not have similar network characteristics as their neighbors regarding degree. It indicates that high-degree keywords do not only bind with other high-degree but they also bind with the low-degree nodes. For instance, the "pain" keyword is the highest-degree node which connects 11,532 other keywords. It links to the other high-degree nodes like chronic pain, opioid, and pain management as well as to 2,393 low-degree nodes like agent-based modeling, ankle surgery, lavender oil, traumatic stress, and sexual assault, whose degree is less than 20.

Not only does the highest-degree node, which is the "pain," associate with other high-degree nodes such as "chronic pain," "back pain," and "opioid," but also with new emerging topics such as "machine learning," "neural network," "regression," and "measurement." The patterns in **Figure 7 (b)** confirm that new topics and connections have emerged in the pain literature over time.

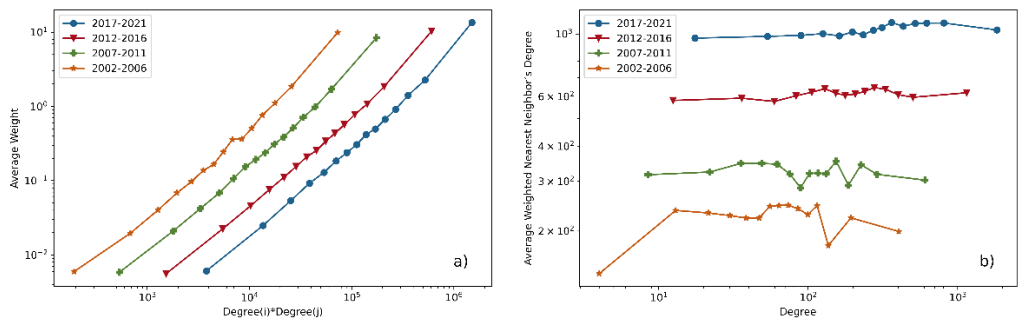

**Figure 7. (a)** Average weight as a function of endpoint degree **(b)** Average weighted nearest neighbor's degree as a function of the degree. Both the x-axis and the y-axis are on the logarithmic scale. The primary takeaway is that high-degree nodes have connections with both other high-degree nodes and low-degree nodes since nodes do not have similar network characteristics as their neighbors in terms of degree.

**Figure 8** shows the weighted clustering coefficient across different values of degree. Over the four time-windows, there is a declining trend in the clustering coefficient, indicating that nodes with a small degree constitute dense clusters more with other small-degree nodes than with high-degree ones. However, high-degree nodes have strong connections with both other high-degree and low-degree nodes. In addition, the clustering coefficient of the 2017-2021 period is always less than that of other time windows. It is evident that in recent years, some nodes have grown as high-degree nodes. For instance: "chronic pain" is one of the highest-degree keywords, and over time, the numbers of keywords connected to "chronic pain" are 303, 738, 2422, and 4611, respectively, in the four time-windows.

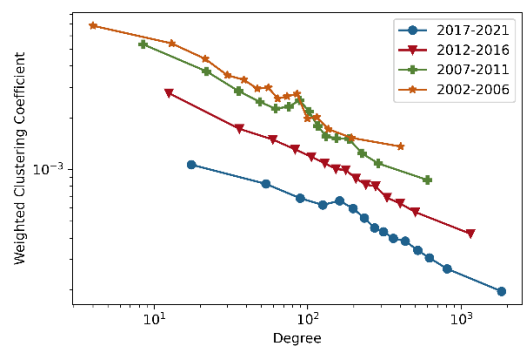

**Figure 8.** Weighted clustering coefficient as a function of degree for four time-windows. Both the x-axis and the y-axis are on the logarithmic scale. The main takeaway is that nodes with smaller degrees constitute more dense clusters with other smaller degree nodes; however, nodes with high-degree have a strong connection with both nodes with high-degree and low-degree.

**Table 4** demonstrates the keywords that have the highest connection density to the neighbor keywords from 2017 to 2021, as measured by the weighted clustering coefficient (See **Eq. 5**). "Clinical decision pathway, noncardiac, intensive critical care, chest pain syndrome, and aha clinical practice guidelines" topics are frequently used with a similar set of keywords in the articles, such as "obesity, sleep, physical activity, lifestyle, chronic pain, diet, cancer survivor, stress, and psychological factor."

**Table 4.** Top 5 keywords that have the highest connection density to the neighbor keywords, as calculated by the weighted clustering coefficient.

| Original Keywords | Clustering Coefficient | Neighbors |
|---|---|---|
| clinical decision pathway | 0.00702 | [obesity, sleep, physical activity, lifestyle, chronic pain, diet, cancer survivor, stress, psychological factor, pain location, pain drawing, pain extent, frozen shoulder, pain sensitivity, questionnaire, experimental pain testing, pain perception, wound] |
| noncardiac | 0.00638 | [obesity, sleep, physical activity, lifestyle, chronic pain, diet, cancer survivor, stress, psychological factor, pain location, pain drawing, pain extent, frozen shoulder, pain sensitivity, questionnaire, experimental pain testing, pain perception, wound] |
| intensive critical care | 0.00614 | [obesity, sleep, physical activity, lifestyle, chronic pain, diet, cancer survivor, stress, psychological factor, pain location, pain drawing, pain extent] |
| chest pain syndrome | 0.00571 | [obesity, sleep, physical activity, lifestyle, chronic pain, diet, cancer survivor, stress, psychological factor, pain location, pain drawing, pain extent, frozen shoulder, pain sensitivity, questionnaire, experimental pain testing, pain perception, wound, breast cancer, pain] |
| aha clinical practice guidelines | 0.00561 | [obesity, sleep, physical activity, lifestyle, chronic pain, diet, cancer survivor, stress, psychological factor, pain location, pain drawing, pain extent, frozen shoulder, pain sensitivity, questionnaire, experimental pain testing, pain perception, wound] |

*4.3. Pain Research Trends*

To further analyze the evolution of pain literature visually, each keyword is placed in one of three following categories for the only purpose of easy visualization:
- Sensors/methods-related keywords (e.g., electromyography, biomarker, machine learning) keywords
- Biomedical-related keywords (e.g., chronic pain, back pain)
- Treatment-related keywords (e.g., surgery, acupuncture)

**Figure 9:** Emerging (left panel) and declining (right panel) keywords in the sensors/methods category from 2002-2006 to 2017-2021. Numbers in parentheses represent the rank and the frequency of keywords, respectively.

**9, 10,** and **11** show the emerging and declining keywords in sensors/methods, biomedical, and treatment categories, respectively. The rankings are assigned based on the keywords' frequency in a specific time window. If a keyword's rank improved from 2002-2006 to 2017-2021 or if a new keyword entered the top 20, then the keyword is considered an emerging topic. On the other hand, if a keyword's rank decreased or if the keyword dropped below the top 20, the keyword is considered a declining topic. The left-side panel of each figure presents the emerging topics, and the right-side panel shows the declining topics. If keywords are in the right-side panel (declining keywords), it is not that they are less important; it only means that researchers are more focused on keywords in the left-side panel (emerging keywords) in recent years.

**Figure 9** represents the trends in the sensors/methods category. Pain management, visual analog scale, random control trial, pain control, electromyography, and pain measurement concepts have been the research focus over the years. Pain assessment and pain intensity topics climbed up in the ranking drastically. Specifically, machine learning and biomarkers have become one of the leading research topics. In 2002-2006, these keywords had never been mentioned in the pain literature; however, during 2017-2021, the machine learning keyword was listed 239 times, and the biomarker keyword was listed 238 times. This change means that researchers have prioritized these topics in recent years. Micro dialysis, magnetoencephalography, electrophysiology, electrical stimuli, and pressure pain threshold subjects were highly used in 2002-2006, but they have not been the main focus of research during 2017-2021 period.

For the past two decades, the research community has investigated pain assessment and pain management approaches. The full potential of pain data was largely untapped in 2002-2006 period because machine learning methods were not actively explored in pain research. Machine learning and statistical modeling techniques have been widely applied to pain assessment and management, starting with the 2012–2016 time-window. Pain researchers explored model building (e.g., deep learning, support vector machines), decision making (e.g., treatment, patient quality of life), and pain measurement (e.g., machine learning, biomarker). However, other modeling methods, such as linear regression, have also been quite common since 2002. We observed that sensors and predictive models have become more common in pain assessment and management.

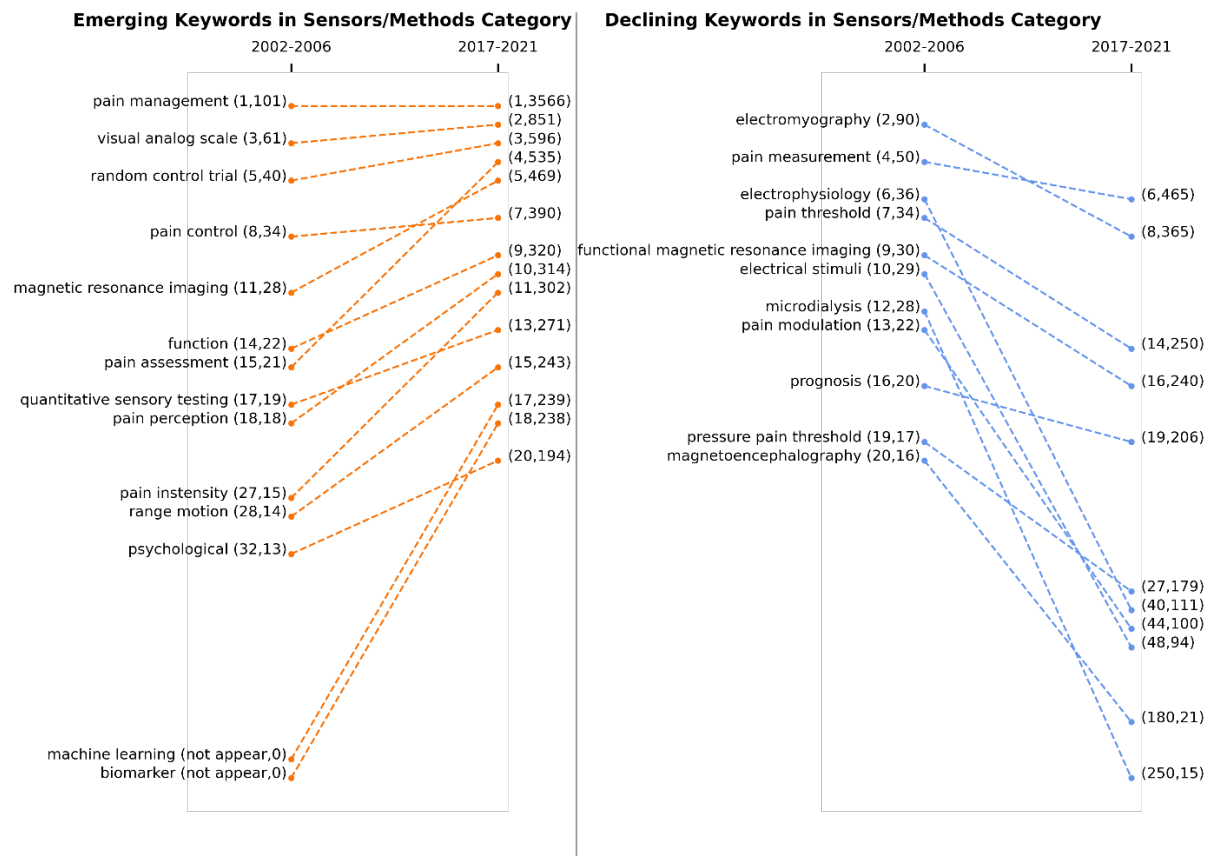

**Figure 9:** Emerging (left panel) and declining (right panel) keywords in the sensors/methods category from 2002-2006 to 2017-2021. Numbers in parentheses represent the rank and the frequency of keywords, respectively.

**Figure 10** shows the trends in the biomedical category. The data reveals that acute pain, musculoskeletal pain, chronic low back pain, cancer pain, and postoperative pain have become increasingly popular over the years and have been actively studied in the last two decades. Meanwhile, chronic pain, neck pain, abdominal pain, low back pain, neuropathic pain, analgesia, back pain, and inflammation have retained their rank on the frequency list and have stayed as the focus of the researchers all through. Dorsal root ganglion, visceral pain, capsaicin, spinal cord, nociceptive, and allodynia have lost popularity over the last two decades.

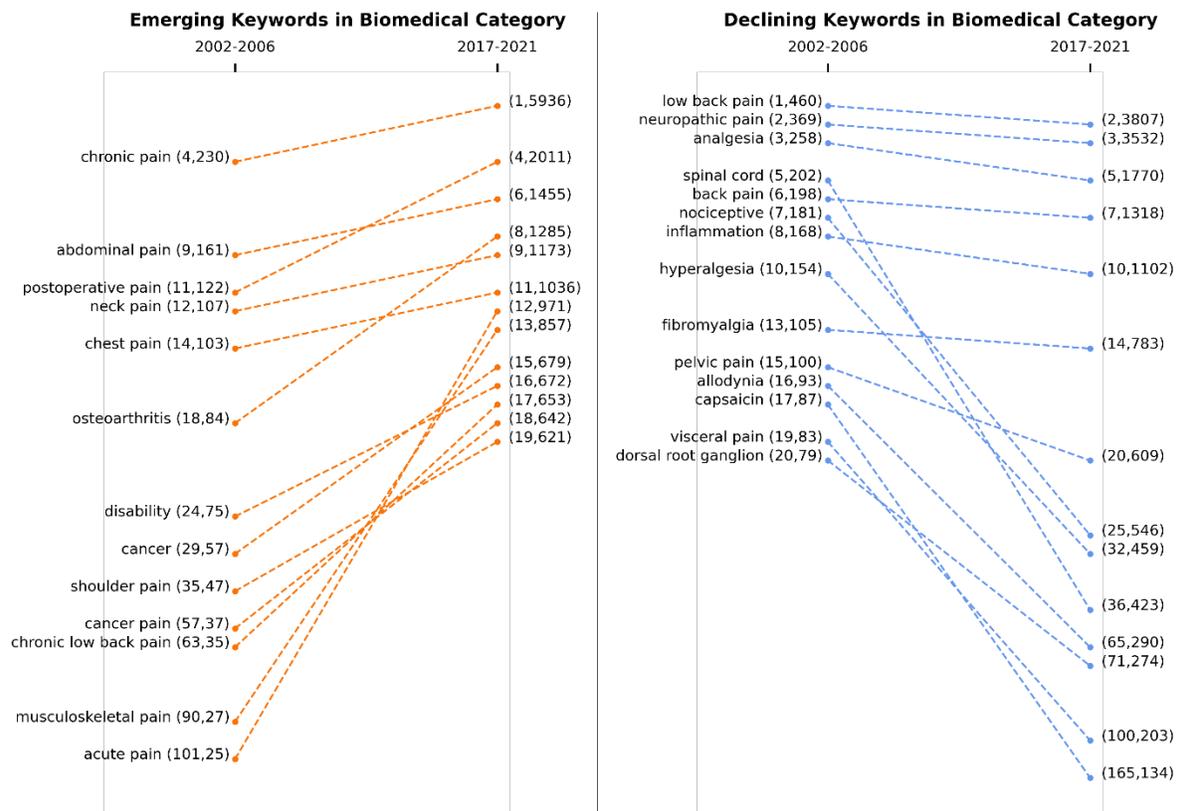

**Figure 10:** Emerging (left panel) and declining (right panel) keywords in the biomedical category from 2002-2006 to 2017-2021. Numbers in parentheses represent the rank and the frequency of keywords, respectively.

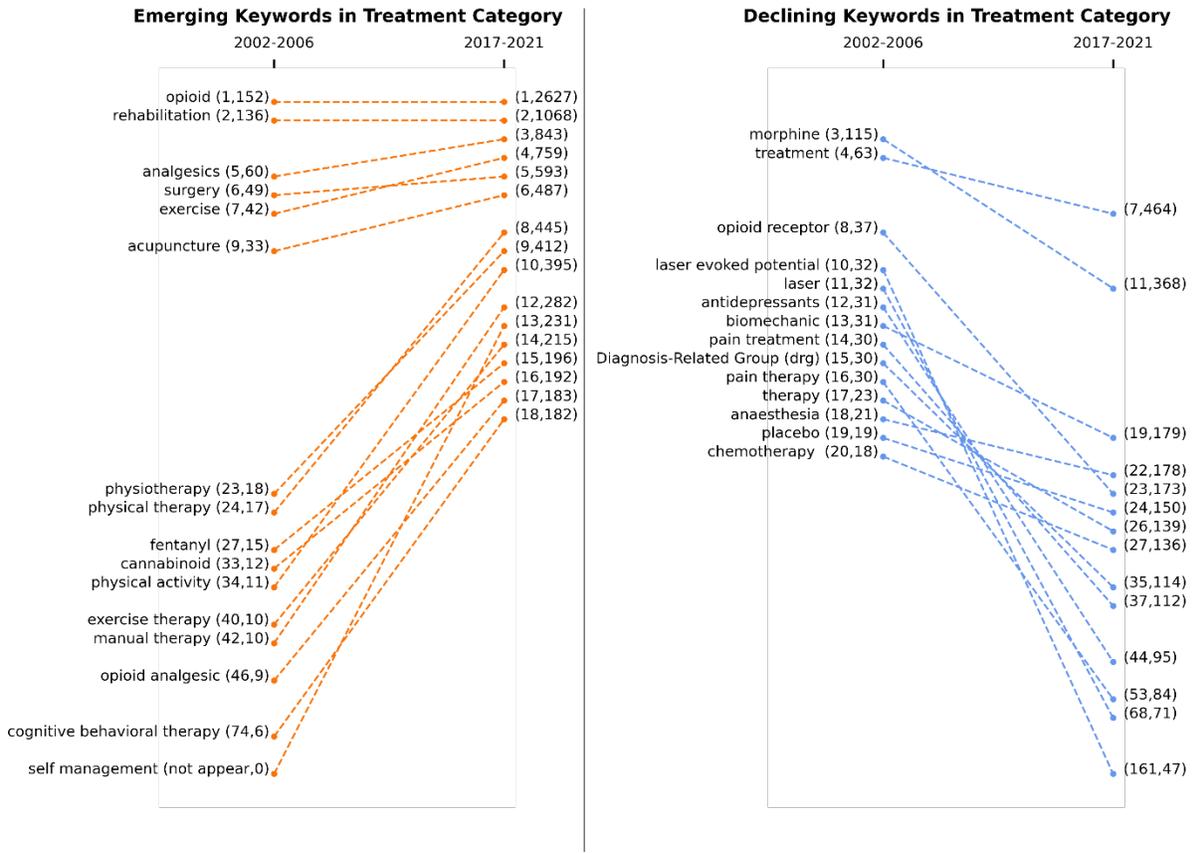

**Figure 11:** Emerging (left panel) and declining (right panel) keywords in the treatment category from 2002-2006 to 2017-2021. Numbers in parentheses represent the rank and the frequency of keywords, respectively.

**11** presents the trends in the treatment category. Opioids, rehabilitation, analgesics, surgery, exercise, acupuncture, and morphine treatments have remained active research topics in recent years. Laser and laser-evoked potential treatment techniques have lost researchers' attention over time. Meanwhile, researchers have directed their interest toward self-management, cognitive behavioral therapy, physical activity, and opioid analgesic topics. The significant changes and trends in keywords since 2002 also reveal the emergence of new technologies for pain management. The early work in the medical field for pain management and assessment has focused on treatment methods such as chemotherapy, but the current research has been emphasizing patient wellness. With the availability of large-scale personal datasets and the development of wearable instruments, the recent work has focused on the quality of life of individual patients, self-management of pain, and plethora of therapy methods.

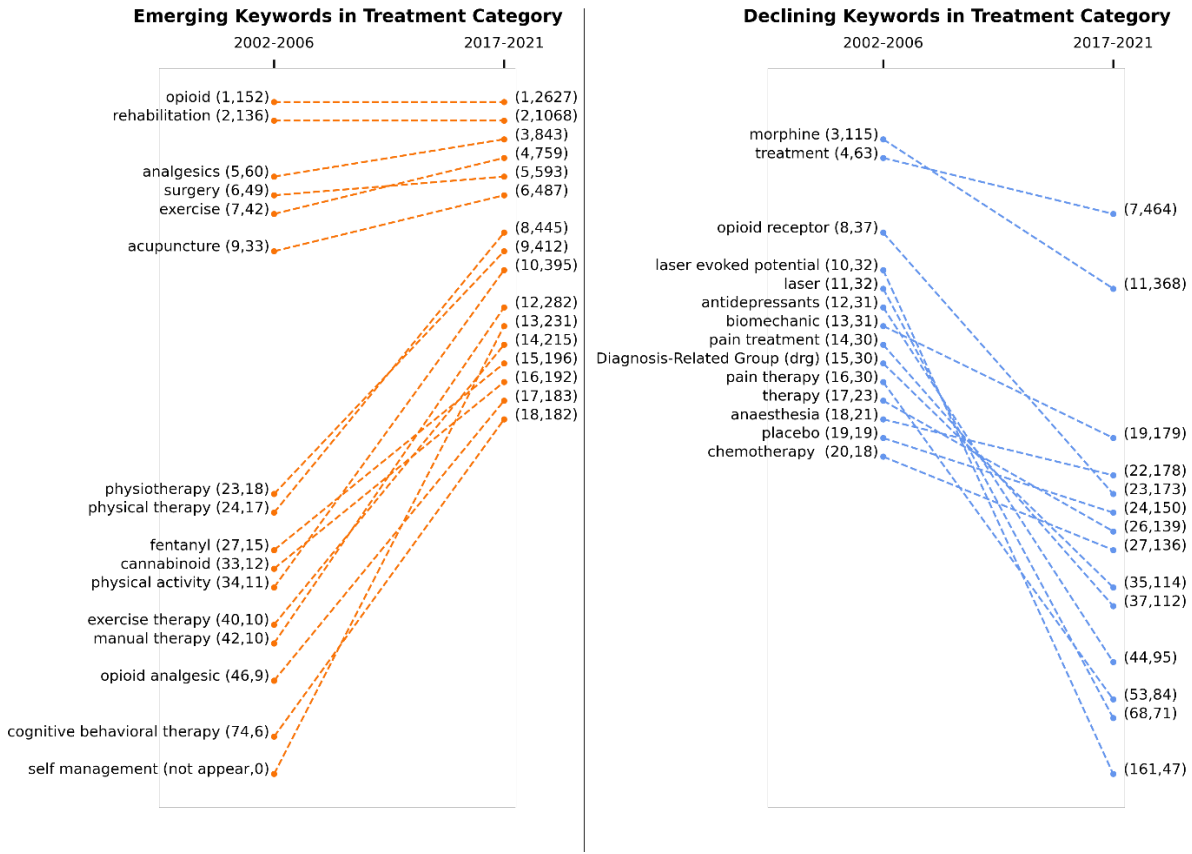

**Figure 11:** Emerging (left panel) and declining (right panel) keywords in the treatment category from 2002-2006 to 2017-2021. Numbers in parentheses represent the rank and the frequency of keywords, respectively.

### *4.4. Association Patterns among Pain-Related Keywords*

We have conducted affinity analysis [51] on pain-related articles. Using the Apriori algorithm [52,53], the affinity analysis first finds all combinations of items (also referred to as itemsets) that occur in a large set of transactions with probabilities greater than a desired threshold. Next, the affinity analysis formulates the rule of co-occurrence of items in the transaction set. Typically, the rules take the form "IF antecedent itemset THEN consequent itemset." The strength of each rule is assessed using several measures, such as support count, confidence, and lift. In **Table 5**, we present these measures for the top 15 rules ranked by their lift. In our study, the support count of a rule is the number of pain-related articles (published from 2002 to 2021) that contain both the antecedent and the consequent keyword(s). The confidence of a rule is the conditional probability of the consequent keyword(s) appearing in the article set, given the presence of antecedent keyword(s) in the article set. The lift of a rule indicates the strength/efficiency of the rule informing the occurrence of the consequent keyword(s). In other words, lift is the ratio of the chance of seeing the consequent keyword(s) in an article if we use the rule to the chance of seeing the consequent keyword(s) without the insights from the rule. With lift greater than 1, we have a greater chance of seeing the consequent keyword(s) in the literature if we know that the antecedent keyword(s) appeared in the literature.

**Table 5.** The association rule found in the keywords listed in the pain literature in the past two decades (2002-2021).

| Antecedent Keyword(s) | Consequent Keyword(s) | Support Count | Confidence | Lift |
|---|---|---|---|---|
| bladder pain syndrome | interstitial cystitis | 281 | 0.78 | 425.53 |
| acute coronary syndrome | chest pain | 327 | 0.56 | 79.07 |
| pain; knee | osteoarthritis | 214 | 0.57 | 69.82 |
| chronic constriction injury | neuropathic pain | 250 | 0.76 | 31.39 |
| postoperative/postop | pain | 537 | 0.85 | 6.76 |
| fatigue | pain | 648 | 0.73 | 5.80 |
| cancer | pain | 763 | 0.66 | 5.28 |
| functional/functionality/function | pain | 348 | 0.66 | 5.27 |
| dementia | pain | 225 | 0.65 | 5.18 |
| parkinson disease | pain | 237 | 0.65 | 5.17 |
| nociception/nociceptive | pain | 772 | 0.61 | 4.87 |
| knee; osteoarthritis | pain | 214 | 0.61 | 4.87 |
| sleep | pain | 382 | 0.61 | 4.86 |
| inflammation | pain | 1126 | 0.56 | 4.44 |
| TRPV | pain | 282 | 0.55 | 4.38 |

We set the minimum support count as 200 and the minimum confidence as 0.55. These settings are determined using computational and practical considerations. The Apriori algorithm yielded 15 association rules that meet the support count and confidence criteria set by the authors (**Table 5**).

To illustrate the results above, consider the first row in **Table 5**. It presents the rule "IF *bladder pain syndrome*, THEN *interstitial cystitis*." It means, if *bladder pain syndrome* appeared in an article, then *interstitial cystitis* will also appear in the article with a confidence of 0.78 and a lift of 425.53. Similarly, the second row in the table leads to the following rule: "IF *acute coronary syndrome,* THEN *chest pain*," with a confidence of 0.56 and a lift of 79.07. These rules inform the researchers on the association among keywords or, in other words, affinity among topics in \ pain-related articles.

## 5. Conclusion

Pain is an important medical issue that affects millions of people every day. Pain-related research has grown extensively across different fields in the last two decades. The vast amount of literature on pain-related research makes the traditional literature review process tedious and impractical. This study, using a KCN approach, provides a macro-level picture of the current pain literature.

In this study, we collected 264,560 articles published between 2002 and 2021 from IEEE, Web of Science, PubMed, or Engineering Village by comprehensively searching pain research articles. We extracted all the keywords from these articles and applied data cleaning and text processing techniques to remove duplicates, tokenize keywords, and extract stem words. From these keywords, we constructed adjacency matrices and weighted adjacency matrices. Using these matrices, we built KCNs and analyzed the network features, such as centrality, affinity, and cohesiveness, to understand the knowledge components, knowledge structure, research trends, and emerging research topics.

Through the KCN-based analysis, we observed that pain literature had grown tremendously in the past two decades. The number of articles and keywords has grown by a factor of 3 and 7, respectively. The number of co-occurrences of keywords has grown at more than twice the speed of keywords growth. We identified the emerging and declining topics in the following categories: (1) sensors/methods, (2) biomedical, and (3) treatment. The Results and Discussion section presented the research trend and insights.

The categorization process of the keywords as sensors/methods, biomedical, and treatment has limitations. This study did classify a keyword into only one of the three categories; it did not consider overlapping membership in the categories. Moreover, the categorization for the purpose of visualization of the emerging and declining trends was performed manually banking on the domain knowledge of the authors. Therefore, the categorization may likely to have subjectivity. Future work will consider classifying a keyword into more than one category and expanding the categories in addition to sensors/methods, biomedical, and treatment. It will automate the classification of a keyword in categories using built-in dictionaries in Python, such as PyMedTermino. In addition, future work will expand the keywords by extracting all words from the abstract. In future work, the KCN-based methods will be extended to analyze the connections between the authors of the articles to reveal potential collaboration patterns across pain-related fields.

**Conflict of interest statement**

The authors have no conflicts of interest to declare.

**Acknowledgements**

This work is partially supported by the National Science Foundation under Grant No. SCH-1838621. Any opinions, findings, and conclusions or recommendations expressed in this material are those of the author(s) and do not necessarily reflect the views of the National Science Foundation.

**Author contributions:**

Conceptualization, B.O. and Z.L.; methodology, B.O. and Z.L.; software, Z.L.; validation, B.O., Z.L. and F.P.; formal analysis, B.O. and Z.L.; investigation, S.K. and F.P.; resources, S.K.; data curation, Z.L.; writing—original draft preparation, B.O.; writing—review and editing, S.K., and Z.L.; visualization, B.O. and Z.L.; supervision, S.K.; project administration, S.K.; funding acquisition, S.K.

REFEERENCES: